# Replay Fidelity in LLM Agent Evaluation: Randomness, Hidden State, and Test Oracles in AppWorld

**You Liu and Quanchao Lu**
Independent Researcher

## Abstract

Reproducibility in large language model (LLM) agent evaluation depends on replay fidelity: which parts of an execution trace are reconstructed, controlled, and compared. This matters for agent debugging, failure analysis, and counterfactual evaluation of stateful tool use. We audit all 71 unsuccessful executions in a fixed historical AppWorld corpus using 284 isolated fixed-code replay attempts. Reconstructed declared settings admit 42 cases under printed-output, pre-oracle database-delta, and native-verdict/test-count checks; resetting the random-number generator before each interaction admits 68. The 26-case gap contracts to four under database/oracle checks, with one endpoint unknown. Mechanism studies identify four materialized-state discrepancies despite matching outputs and verdicts. In one admitted endpoint pair, an identical API call yields matching responses but different row effects; common future Python RNG removes that deterministic difference in two paired repetitions. A separate JWT credential-equality program and alignment controls expose authentication/write differences, with missing repetitions retained. For all 10 affected post-oracle export cases, native export from retained output trackers restores the reference comparison without changing original cached contents. A purposeful 12-case BFCL pilot shows no discrepancy under its tested public-state, return, and verdict checks. We preserve resource censoring and infrastructure deviations. The contribution is a bounded empirical characterization of replay eligibility, hidden state, and test-oracle integration, without estimating debugger accuracy, repair gains, or method-ranking distortion.

Keywords: LLM agents; agent evaluation; reproducibility; record-and-replay; replay fidelity; hidden state; test oracles; agent debugging; AppWorld; BFCL.

## 1 Introduction

Agent debugging studies reuse failed executions to inspect a mistake, reconstruct a prefix, or test a correction. Each use needs evidence about what was reproduced. Matching printed observations, database changes, a task checker, and the state needed by a later program are related but different requirements. AppWorld supplies a concrete setting in which these distinctions can be inspected through native applications and programmatic evaluation [1]. A replay rejection can reflect output rendering, an unobserved endpoint, or a validator reading another state view; an acceptance can leave relevant execution state unspecified.

We examine a fixed historical failure population under two reconstruction controls and several explicitly named validation targets. Our central contrast keeps source cases fixed. The large strict-admission difference is mostly absent at the checked database/oracle endpoint projection. This negative result limits an interpretation of every strict text rejection as failed endpoint reconstruction. Conversely, greater admission under a weaker projection does not establish that a reconstruction supports every later operation.

We ask three research questions. RQ 1: How does execution control interact with the validation target in admitting this cohort? RQ 2: Which differences can materialized rows or specified continuations expose when checked projections agree? RQ 3: Can post-oracle exclusions be localized to the export reference and corrected without changing original output rows?

Agent failure attribution and debugging study where errors occur and whether interventions improve outcomes [13-17]. Controlled replay, restoration checks, paired repair baselines, and provenance also have substantial precedents [2-7,11,12]. Our contribution is a crossed finite-cohort comparison of reconstruction control and validation target, with materialized-state, credential/RNG, and export-reference mechanism tests. It does not introduce replay certification or estimate new debugging-method performance.

### 1.1 Contributions and scope

First, we enumerate admission outcomes for all 71 historical native-evaluator failures and show how the N/C gap changes from 26 cases under printed-output/database/oracle agreement to 4 under database/oracle agreement. Second, we materialize four database discrepancies, distinguish one accepted endpoint pair with specified native continuations and controls, and validate a source-directed export correction on every affected C case. Third, we provide locally frozen provenance that retains unknowns, adapter corrections and infrastructure deviations. A bounded BFCL check supplies a consistency boundary, not another natural counterexample.

The study concerns one old AppWorld runtime and source harness. It measures neither live diagnosis nor a matrix of debugger outcomes. The claim is what these checks establish for these reconstructed executions and declared programs. Modern runtimes, other policy lineages and broader prevalence are unmeasured.

## 2 Study design and evidence objects

*Table 1. Evidence units and execution accounting. Rows overlap in source cases and must not be pooled into a failure sample size.*

| Study block | Source unit | Attempts | Role/retained missingness |
|---|---|---|---|
| Primary AppWorld census | 71 failures, 32 families | 284 | N/C x 2 fresh processes; one C resource-censored case |
| Fixed-code completion | 25 early N text stops | 50 | 24 endpoints resolved; one unknown in both repetitions |
| Earlier mechanism studies | One credential case; 10 export cases | 16 + 20 | Interpreted programs and binding-view probes; kept separate |
| New closeout supplement | Existing cases and known control | 66 | 58 complete observations; 8 censored/unknown |
| Independent BFCL pilot | 10 failures + 2 successes | 24 | All tested contracts match; not a random prevalence sample |
| Separate development/control records | Not new source failures | Separate | 8 closeout preflights; BFCL gate/preflight; census controls retained |

### 2.1 Corpus and reconstruction

The source inventory contains 147 official historical train/dev ReAct executions generated under the GPT-4o-2024-05-13 alias: 76 native-evaluator successes and 71 failures, with 32 failure-bearing task families. The study includes every failure. Thirty earlier-inspected failures and 41 previously unmeasured failures were rechecked with one primary runner. The output bundle is version 0.1.3; per-record code, data and evaluation metadata are 0.1.0. The reference endpoint is the archived agent's final database change stream, not a gold correct-solution terminal state. No held-out test-task contents enter the analysis.

Each reconstruction starts a fresh AppWorld 0.1.0 world and executes the archived code unchanged. N uses reconstructed declared settings: seed 123 and a 60 s cell timeout. C additionally calls random.seed(123) immediately before every recorded interaction, uniformly across cases. This is a controlled execution-policy change, not identification of the historical controller's RNG lifecycle. Expected outputs do not select the seed. The primary protocol stops at the first output mismatch; a separate fixed-code completion study follows the remaining recorded code without replacing actions.

### 2.2 Admission and three-valued evidence

Let O denote equality of all printed outputs after only CRLF normalization and outer-whitespace stripping. D_pre denotes byte equality of all 12 archived database delta files with both immutable automatic and explicit pre-oracle snapshots. V compares the native success flag and passed/failed/total test counts, not individual test identities. Primary admission requires O AND D_pre AND V in both fresh processes. It does not check every API argument, Python variable, RNG end state or private service object.

We retain match, difference and unknown separately. A missing endpoint is not an observed state difference. A conjunction can still be known false when one checked output differs even if its final state is unmeasured. In particular, N's text rejection and its unknown completion endpoint are compatible statements. Two fixed-seed processes assess technical repeatability here; they do not estimate a stochastic failure probability.

### 2.3 Isolation, budgets and deviations

Primary attempts use network-disabled containers, read-only inputs, two CPUs, 4 GB RAM, at most two concurrent tasks, and a 120 s external watchdog with 5 s kill grace. Twenty sentinel attempts are included in 284, not added to it. The closeout uses 180 s ordinary caps and a single predeclared 600 s resource supplement, retaining the 60 s cell limit. Its two infrastructure incidents censored four attempts and defeated timely cap enforcement. The host wrapper was corrected and all partial records retained; the scientific worker did not change. An expired 3 h scheduling window was amended only to finish the original six pending jobs inside the unchanged 4 h overall bound. These deviations are part of the evidence, not erased by successful integrity checks.

All reported replays use existing traces and local CPU, with zero new model API calls. Historical generating rollouts and excluded legacy paid development are different evidence units. Manuscript preparation starts from a hash-verified read-only archive and invokes no new task execution.

# 3 Admission depends on the validation target

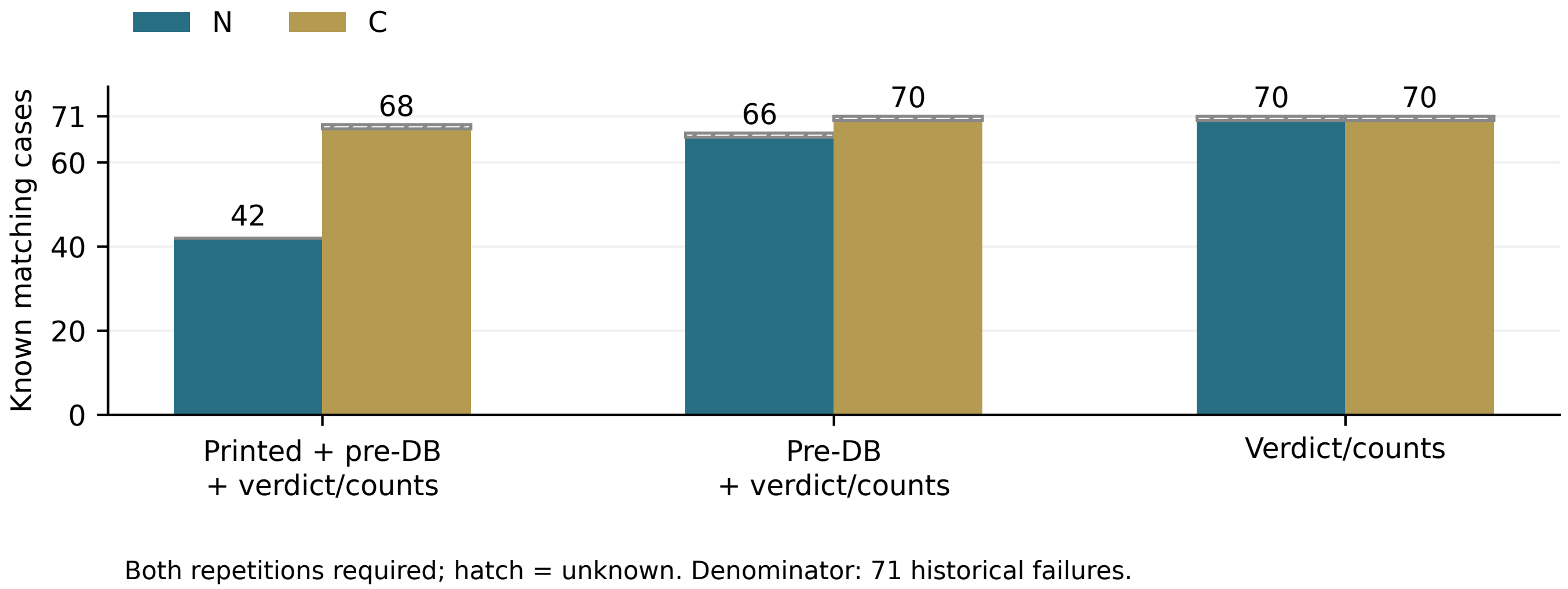

*Figure 1. Known-match counts on the same 71 failures. Hatched segments are unknown components/endpoints, not failures. The known-match N/C gap is 26 under the primary rule, 4 under D_pre AND V, and 0 under V; one endpoint remains unmeasured under both controls for the latter comparisons.*

## 3.1 Primary admission and population composition

N admits 42/71 cases (59.2%); C admits 68/71 (95.8%). Forty-two pass both controls, 26 pass only C, none pass only N, and three are not admitted by either. The last group contains two C text differences and one C resource-censored case, rather than three proven semantic restoration failures. Admission decisions agree across both repetitions for every primary case/control pair.

The earlier 30 failures remain N = 20/C = 28; the remaining 41 contribute N = 22/C = 40. Admitted families increase from 21/32 to 30/32. Family macro admission is 58.33% under N and 92.71% under C. Removing any one family leaves the C-minus-N admission difference between 33.82 and 38.24 percentage points. This is sensitivity within a fully enumerated historical cohort, not a confidence interval for future agents.

*Table 2. Contract ablation: matched/different/unknown cases. Both process repetitions must support a match.*

| Validation target | N | C |
|---|---|---|
| O AND D_pre AND V | 42/29/0 | 68/2/1 |
| Exception-normalized O, D_pre, V | 42/29/0 | 70/0/1 |
| D_pre AND V | 66/4/1 | 70/0/1 |
| V only | 70/0/1 | 70/0/1 |

## 3.2 Resolving early-stop missingness

The 25 first-output N stops are token-expiry text variations. Executing all of their remaining archived code, twice, resolves 24 checked endpoints that match D_pre and V; the remaining endpoint is censored. The four other N exclusions already complete with matching outputs and V but different pre-oracle deltas. C's two exception-rendering differences are resolved by separate diagnostics without changing the primary decisions. They concern an extra list-comprehension frame and a NameError suggestion.

Consequently, simply calling every text stop an endpoint failure would overstate the measured reconstruction gap. Equally, masking text is not a demonstrated solution for live agents: an agent could react to a changed observation. The ablation answers a fixed-recorded-code endpoint question. It neither re-samples an actor nor proves that different printed histories induce the same policy. DFAH already distinguishes comparison levels for repeated agent behavior [12]; the present result quantifies their interaction with N/C admission on one fixed failed-execution cohort.

## 3.3 Workload associations are not independent causes

For traces with at least two login requests, admission changes from N = 1/30 to C = 27/30; traces with at most one login pass 41/41 under both controls. Long traces (at least 20 interactions) change from 4/24 to 21/24, and multi-app cases from 0/20 to 18/20. Every multi-app trace in the inventory also has repeated login requests. Counts include failed/repeated calls, so these correlated proxies do not identify independent complexity effects or diagnostic difficulty. The controlled policy contrast establishes conditional admission sensitivity, not selection-induced repair bias.

## 4 Are database differences merely serialization?

*Table 3. Materialized verification of all four N delta-discrepancy cases. Each policy has two fresh-process repetitions; counts are cases, not independent discoveries.*

| Source case | Outputs and V | D_pre N/C | Materialized phone rows | Within-policy repeats |
|---|---|---|---|---|
| 302c169_1 | Match/match | Different/match | Different | Equal, N and C |
| 302c169_2 | Match/match | Different/match | Different | Equal, N and C |
| 3c13f5a_1 | Match/match | Different/match | Different | Equal, N and C |
| 3c13f5a_3 | Match/match | Different/match | Different | Equal, N and C |

### 4.1 Competing explanations and procedure

A change-stream byte mismatch does not alone imply different endpoint contents. SQL order, redundant updates or serialization could differ while the final rows agree. A second risk is querying a different database after evaluation. We therefore retain all 12 original output-cache SQLite connections before an in-process oracle can rebind global model classes. For every table, we sort encoded rows and hash the resulting representation. We independently hash the password-reset fields in phone and Spotify users. These are content projections, not a serialized Python heap or a proof of schema/physical-file identity.

The 16 new attempts cover four existing cases, two policies and two processes per policy. N delta non-matching is the object being measured, not a reason to drop a case. The native verdict and counts are obtained in a separate child process. Parent output-row fingerprints, bindings, RNG state and persisted deltas are checked across that evaluation. No expected-state file is used to construct execution state.

### 4.2 Results

All four pairs differ in actual phone table-row fingerprints under N versus C in both repetitions. Differences include users, global_text_messages and associated text-search tables; the directly queried reset-code field hashes also differ. Each policy reproduces its own materialized fingerprints across repetitions. C matches the archived 12 delta files and N does not. The cases belong to two task families, not four independent environments.

These measurements reject a purely change-stream-representation explanation for the observed N/C discrepancies. They do not establish that every column difference is independently consequential. We do not infer exclusive per-column causes from positionally aligned SQL operations: such an auxiliary trace diagnostic is retained but is not used as a materialized per-column proof. Entity-row hashes and direct reset-field queries supply the stated evidence.

### 4.3 What this does and does not say about the oracle

The native success flag and aggregate test counts still match the source. This is not evidence that the benchmark's task oracle graded incorrectly. An oracle checks specified requirements rather than every persistent field. Here, verdict/count matching and exact endpoint-state matching answer different questions. All source cases in this comparison are unsuccessful executions; keeping them unsuccessful is also weaker than reproducing all of their context.

The four cases are admitted by O AND V but excluded by the stronger primary O AND D_pre AND V rule. They must not be presented as accepted-primary-contract counterexamples. Their role is to show that the four-case residual gap in Table 2 has materialized data evidence, rather than only a serializer artifact. Whether those data differences change a relevant repair's result remains unmeasured.

### 4.4 Boundary established

A useful statement of replay fidelity names the database object, observation stage and comparison semantics. Neither exact delta bytes nor an aggregate score is universally preferable. Exact deltas can be stricter than semantic equality; aggregate scores can miss data needed by another operation. The measured distinction motivates the next experiment: two reconstructions whose checked rows do agree, followed by a specified continuation.

# 5 Identical API arguments, omitted RNG state

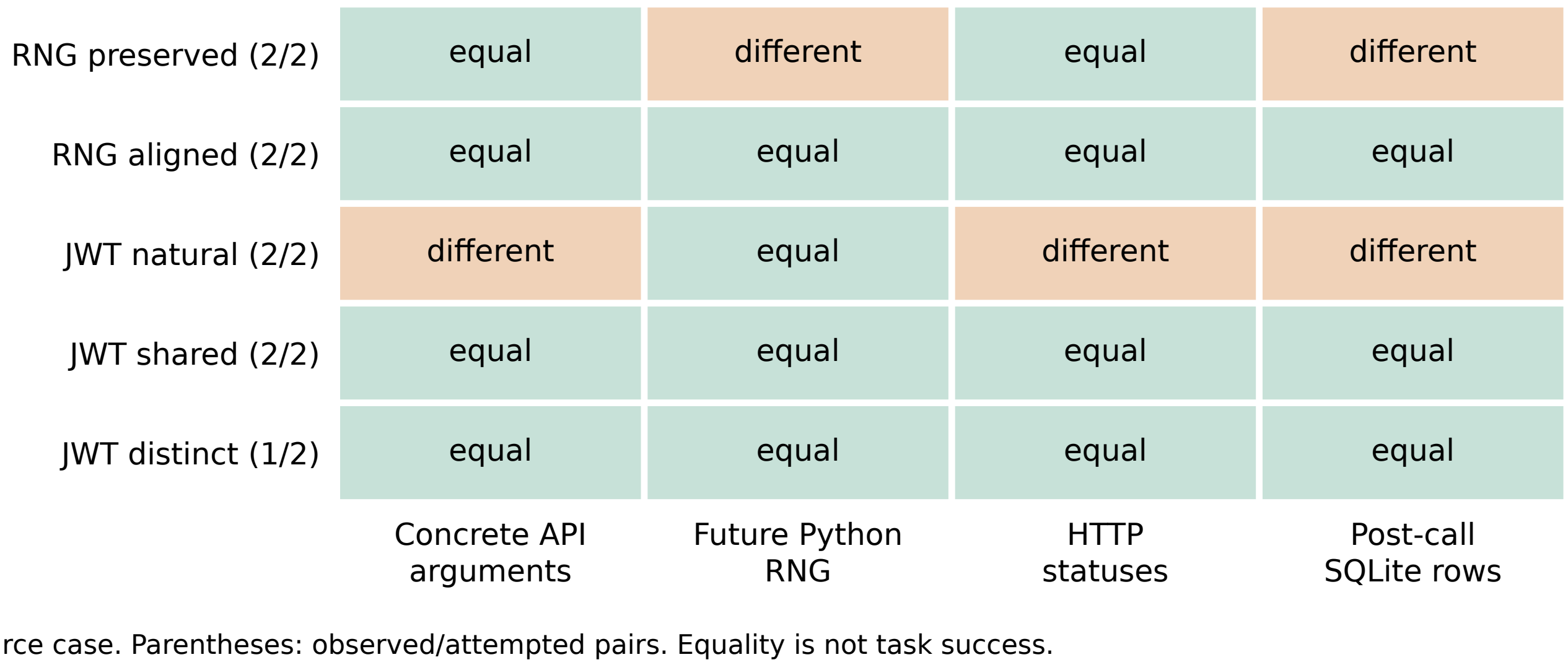

*Figure 2. One source case; equality between N/C for each added program. Parentheses show observed paired repetitions out of two attempted pairs. All programs are fixed. The natural JWT program resolves a different logout argument; one distinct-token pair is infrastructure-unknown. Colors denote equality, not task success. Prefix canonical API-argument matching is outside the baseline qualification rule.*

## 5.1 Accepted endpoint pair and literal probe

The source case e85d92a_2 has 65 recorded interactions, 20 login calls and 1,365 API requests. N and C both satisfy the original printed-output, 12 pre-oracle delta and verdict/count checks. Their materialized table-row fingerprints also agree before the probe, while their Python RNG states differ. The baseline oracle runs in a child process and its non-interference with the measured parent projections is verified.

We execute one public-wrapper statement through AppWorld.execute: send_password_reset_code with the same predeclared literal email constant and raise_on_failure=False. The manuscript omits the synthetic address, but its concrete argument hash is identical across N/C. The target does not consume the current-token variable. Task time stays fixed; no clock-changing permission is added.

## 5.2 Preservation and alignment contrast

With each reconstructed endpoint's RNG state preserved, both policies return HTTP 200 and the same printed response, yet their post-call Spotify user and Gmail entity-row fingerprints differ. Reset-field hashes differ as well. This occurs in both paired repetitions. Pre-call rows, concrete arguments, program bytes and measured response equality rule out those observed objects as the source of the disagreement.

A fixed control aligns only future Python RNG by random.seed(20261107) before that same statement. Credential variables remain as reconstructed. The post-call row difference disappears in both pairs. This supports an RNG-dependent deterministic-state explanation for this declared API under these controls. It does not establish equality of every hidden object or a sufficient contract for all programs.

## 5.3 Scientific meaning

Omitted RNG-stream position is an established replay hazard, measured for agent snapshots by Pakhomov and Nijkamp [11]. This native API experiment is a scoped AppWorld extension. Different reset codes can both be valid: the result concerns deterministic row-level distinguishability, not unequal outcome distributions, task success, repair benefit, or harm. It does not make per-interaction reseeding a generally correct restoration policy. Section 6 tests a separate credential-value relation with an authentication/write effect under common future RNG.

## 6 JWT equality and actual program alignment controls

```
# Schematic public-wrapper listing; symbolic values shown.
# Result assignments and printing are omitted here.
apis.spotify.logout(access_token=current, raise_on_failure=False)
apis.spotify.show_account(access_token=shared, raise_on_failure=False)
apis.spotify.create_playlist(title=FIXED_TITLE, is_public=False,
                            access_token=shared, raise_on_failure=False)
```

*Table 4. Native public-wrapper program. Read/write HTTP results concern the shared token; logout returns 200 in all observed arms. Missing pairs are not imputed.*

| Current binding | Observed pairs | Shared read/write N | Shared read/write C | Playlist rows N/C |
|---|---|---|---|---|
| Natural | 2/2 | 200/200 | 401/401 | +1/0 |
| Forced shared | 2/2 | 401/401 | 401/401 | 0/0 |
| Forced distinct | 1/2 | 200/200 | 200/200 | +1 /+1 |

### 6.1 Relationship rather than a token-string anomaly

Only the first JWT is printed by this source program. Its historical and C logs contain one distinct token; N uses 20. This is the sole N-admitted repeated-login case in the census: the printed projection exposes only part of its credential history. At the endpoint, the current credential equals the initially visible shared token under C and differs from it under N. Token strings differing alone would not establish a semantic consequence. We test a program whose meaning depends on that equality relation.

The pinned login code samples a JWT expiry; repeated generation at the fixed task time can produce equal signed values under the controlled RNG. Logout blacklists the supplied token value in a manager set. Database deltas therefore do not encode all of the relevant credential state. This is value equality, not a claim about Python object-address identity or a vulnerability in a live authentication service.

### 6.2 Native execution and causal controls

Earlier mechanism records interpreted endpoint programs. The 12-attempt closeout instead executes the actual public-wrapper Python program through AppWorld.execute, instrumenting the native request delegate without replacing arguments or responses. After qualifying the source endpoint, it injects the same shared token, uses a common suffix seed 20261107, and leaves task time unchanged. In the natural arm, current is the source variable. In the other arms, current is explicitly bound to shared or to the fixed distinct token obtained from a recorded N endpoint. These are controlled synthetic perturbations, not additional natural cases.

The natural program reproduces N = 200/C = 401 for the shared read and write in two process pairs, with playlist increments+1/0. Forcing the shared binding makes both policies reject those calls in both pairs. Forcing the distinct binding makes both accept them and produce equal post-call rows in the one observed pair. Its second N/C pair was infrastructure-censored during prefix replay, before qualification and before the binding change; it stays missing. Ten of 12 attempts completed observations.

### 6.3 Intervention-class and fidelity boundaries

The natural arm is a same-program witness: the preceding logout resolves different concrete token arguments. Only the later shared-token read/write arguments match. The literal RNG experiment in Section 5 is the separate same-concrete-argument witness. Neither is a new live-agent continuation or an original-task repair, and we do not pool their effects into a repair success rate.

Existing prefix request logs already differ in their token arguments. A stronger canonical-request matcher could reject this pair; we have not run or falsified SymTrace's represented-state guarantee [3]. The conclusion concerns the declared output/delta/verdict admission rule. The historical controller's complete RNG state is not identified. Earlier expiry-grid probes advance time outside the default allow_datetime_change=False contract and remain boundary records, not headline evidence.

## 7 Correcting the post-oracle export reference

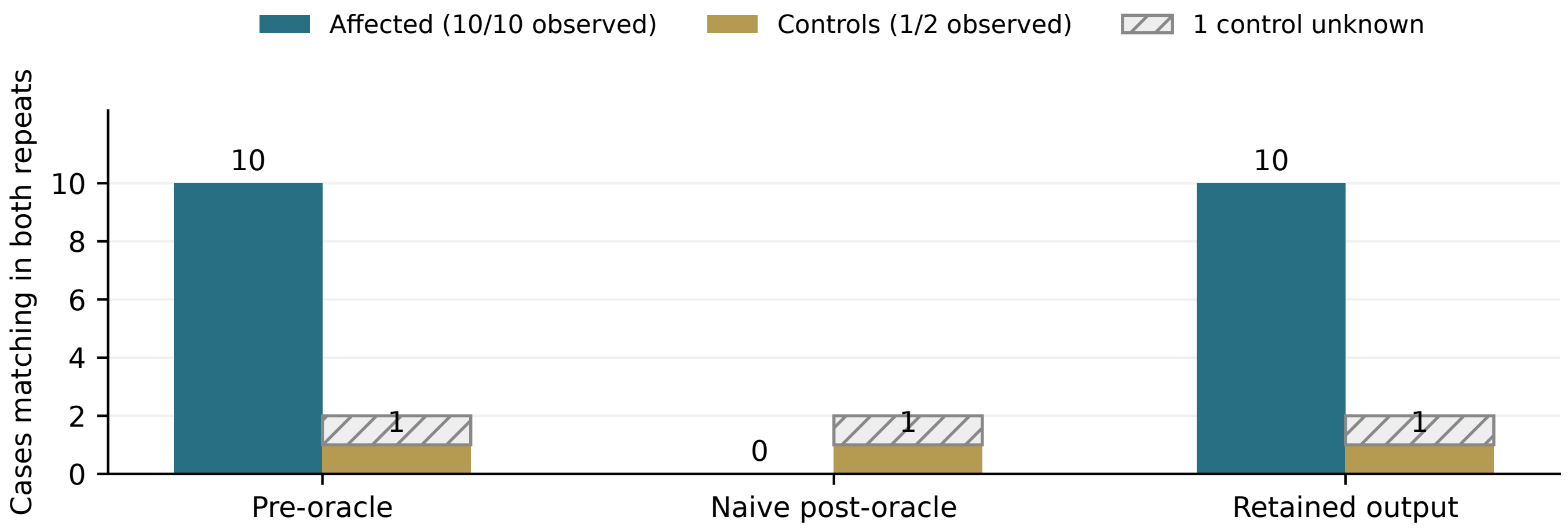

*Figure 3. All 10 affected C cases and two fixed unaffected controls were attempted twice. Ten affected cases and one control completed both repetitions; the other control is hatched as unknown. Retaining the original output trackers corrects the export comparison without changing their contents or the original cached table rows. This is an export-source correction, not restoration of the Python heap or global bindings.*

### 7.1 Observation order and competing explanations

Replacing the primary pre-oracle delta check with a naïve post-oracle export would reduce admission from N = 42 to 35 and C = 68 to 58. This is a stage/reference mismatch, not simply a different equally appropriate contract. The ten affected C cases span four families. Earlier fresh-process probes on every one of them retain output engines and trackers before evaluation. All preserve original table-row and tracker hashes while affected ORM class bindings change to input-database homes.

The app-specific affected export files involve phone in three cases, venmo in six and supervisor in one. The pinned export helper serializes the currently bound class database. A home/binding change alone does not guarantee different exported bytes: the corresponding trackers can be equal. One case changes additional bindings while only the supervisor export differs. We therefore compare actual source objects and files instead of using a binding-change flag as an exclusion detector.

### 7.2 Source-directed mitigation

The new 24-attempt experiment fixes all 10 affected cases plus two short unaffected controls before execution. It retains original output trackers by their output-cache identity before the in-process oracle, then invokes each retained tracker's native save operation. Expected delta files are read only afterward for validation; the writer neither copies gold bytes nor uses them to select data. Content, tracker and binding probes check the scope of the action.

Every affected case completes twice: the naïve export mismatches, the retained-output export matches all 12 archived delta files, and original table rows and trackers remain unchanged. One unaffected control also completes twice and keeps matching. Both attempts for the other control are infrastructure-unknown before qualification, so the control result is 1/2 cases observed, not a zero-false-positive estimate. Overall 22 of 24 attempts complete observations.

### 7.3 What was corrected

The mitigation corrects what the validator exports. It does not undo global ORM rebinding, serialize variables, or demonstrate safe arbitrary continuation after an in-process oracle. When later execution is required, the independently measured child-oracle path avoids altering the measured parent projections. An immutable pre-oracle snapshot also avoids this comparison-stage error.

The evidence does not show that the original output database was destroyed or that AppWorld's returned task grade was wrong. It localizes a concrete integration mechanism in version 0.1.0. Oracle state integrity has close prior work [7]; the contribution here is the complete affected-case verification and a mitigation with a measured source-reference boundary, not the first observation that evaluators can affect state.

## 8 Independent BFCL check and retained uncertainty

*Table 5. Separate evidence blocks. Closeout totals are attempts; BFCL pilot totals are source cases qualified in both repetitions.*

| Block | Planned | Complete observations | Censored/unknown |
|---|---|---|---|
| Literal RNG continuation | 8 attempts | 8 | 0 |
| Actual JWT program | 12 attempts | 10 | 2 infrastructure |
| Materialized DB checks | 16 attempts | 16 | 0 |
| Oracle export correction | 24 attempts | 22 | 2 infrastructure |
| 600 s endpoint supplement | 4 attempts | 0 | 4 resource-censored |
| Known BFCL cwd calibration | 2 attempts | 2 | 0 |
| Closeout total | 66 attempts | 58 | 8 |
| Separate natural BFCL pilot | 12 cases /24 attempts | 12 cases | 0 measured discrepancies/unknowns |

### 8.1 Bounded second-environment consistency result

BFCL is an independent native function-calling evaluation backend [8]. We pin bfcl-eval 2025.12.17 and code/data commit f7cf7359b7ac615a0b294831c5ba2bc95ee4a000; 181 packaged files match that commit's Git blob hashes. Recorded outputs come from result commit cadd76a8dc62984e7bf15ecaf358f070bae745cc, batch 2025-12-16, Qwen_Qwen3-8B-FC, multi_turn_base. This is a historical replay check, not a new model score.

The frozen 12 IDs are 0, 1, 2, 3, 4, 5, 6, 9, 51, 52, 106, 152; success controls are 1 and 52. They cover seven stateful tool classes. Class-coverage selection is purposeful, not random. Each case runs twice in a fresh network-disabled process using native decoding, execution and checking; pure native function bodies are loaded without constructing an inference client. All tested decoder, exact-return, JSON-return, logged public-state, native-outcome and diagnostic checks match. The 24 attempts compare 254 decode events, 196 tool returns and 114 public snapshot groups. Every source failure remains a failure; both successes remain successes.

No observed public-state change occurs across the native oracle; checked Python RNG and decimal contexts also remain unchanged. This does not establish complete private-state equality. Calls are heavily weighted toward the file system (73 of 98 calls per repetition). The small coverage pilot cannot estimate a population discrepancy rate or show the same AppWorld mechanism transfers.

### 8.2 Known control rather than a new counterexample

A separate constructed native file-system calibration compares full deepcopy with an explicitly weak public-root-only reconstruction. Both pass the native public-state check before the same touch action. The full copy writes in the original child directory; the weak copy writes at root. Both repetitions reproduce this boundary. CAREFUL-Bench already identifies the omitted private current directory and later-call exposure [9]. We count this as observer calibration, not a second natural discovery.

### 8.3 Missingness remains in the record

The four 600 s attempts on the sole unresolved AppWorld endpoint complete 84, 84, 91 and 91 of 99 interactions before normal external termination. The final database/oracle endpoint stays unknown. N's observed text mismatches already falsify strict text equality; unknown here concerns the unmeasured endpoint. No further time extension or case substitution is made.

The other four closeout unknowns arise from two infrastructure incidents. Cleanup did not promptly stop the workers; the first host wrapper raised before recording its result, and its two host records were reconstructed from preserved launch/partial logs. The corrected wrapper recorded the second incident and stopped scheduling. Causes of large wall/monotonic gaps remain unconfirmed. Cleanup, a bounded keep-awake guard and the original-inventory window amendment are disclosed. They do not become benchmark failures or successful repetitions.

# 9 Interpretation, validity and practical implications

## 9.1 The interaction, not a universal replay rate

The central empirical result is the interaction between a reconstruction controller and the object being validated. A 26-case strict gap is largely absent at the checked fixed-code database/oracle endpoint, leaving four materialized discrepancies and one unmeasured endpoint. This narrows what an admission difference can mean. It does not make the weaker projection preferable: the continuation studies supply concrete state relationships that the original checks omit.

Record/replay already requires explicit treatment of nondeterminism [10]. Agent work has measured RNG-stream omissions and emphasized restoration checks [11], while DFAH separates comparisons of decisions and tool paths [12]. Our empirical distinction is the N/C-by-target interaction and its concrete state/reference mechanisms in archived AppWorld failures. The study does not claim the first observation that hidden state or comparison level matters.

## 9.2 Deterministic effects and repair utility

An admitted projection is evidence about the objects it compares. For one declared program, two accepted reconstructions can yield different observations or rows; this does not estimate how often that happens, or identify a sufficient representation for every program. In particular, the literal RNG probe may generate two equally valid codes. Its deterministic divergence does not prove unequal outcome distributions. The credential program has a real authentication/write difference, but it is not a repair of the source task.

Selection-induced debugger bias needs additional outcomes. If g_m (i) is method m's gain over a common continuation baseline on case i, admission changes which g_m values would be averaged. A changed comparison further requires different relative method effects across retained/excluded strata. Our census observes admission, not this outcome matrix. Missing repair effects cannot be filled by feature weighting, and absence of ranking evidence is not a falsification of possible ranking distortion.

## 9.3 Threats to validity

Construct validity: exact delta bytes, table-row fingerprints, printed text, verdict/counts and program responses are intentionally distinct targets. We do not compare individual test identities or certify a serialized heap. Full-trace admission differs from recovery at an earlier debugger anchor. Source agent-end deltas differ from a correct-solution reference state.

Internal validity: the original Python/controller stack is incompletely identified; C is a uniform intervention, not a proven reconstruction of that lifecycle. Parent-oracle isolation is checked on measured projections, not every object. Output mismatch completion holds recorded code fixed, so it cannot predict a live actor's response to changed observations. Host workloads are uncontrolled; four infrastructure cap violations and eight closeout unknowns limit those specific contrasts. The final scheduling amendment and the one-observed-pair control are visible rather than averaged away.

External validity: the main cohort has one old AppWorld version, one source harness/model lineage, 32 failure-bearing families and no new policy sampling. Cases in a family are related. Mechanism cases are selected for explanatory value; one behavioral case and a constructed known control do not estimate prevalence. BFCL's 12 purposeful cases and different state representation are a boundary check, not proof of the same causal mechanism across benchmarks. Broad production prevalence is not part of the central claim.

Conclusion validity: counts describe this finite cohort. Two technical repetitions are not independent agents or a calibrated failure probability. We report family sensitivity and available denominators rather than infer future-population confidence from calls. Integrity checks passing does not imply all execution protocols were followed: cap conformance is explicitly false on four closeout attempts.

## 9.4 Guidance grounded in the observations

Name the reference object and observation stage; separate attempt completeness from endpoint knowledge; report a contract ladder rather than one context-free replayability flag. Keep immutable pre-oracle execution snapshots. If exporting after evaluation, verify the data source rather than trusting a mutable class binding. Declare the continuation program class and randomness coupling; use fixed-argument and state-alignment controls when explaining a behavioral witness. Retain missingness and complete the same frozen case inventory without searching for replacement successes.

The current evidence supports empirical characterization and scoped mechanism/mitigation claims. A future request-level admission audit could test stronger validation targets on archived logs; a common-condition repair comparison would be needed to test utility or ranking effects. Such extensions need a separate protocol. The present experiment results remain frozen.

# 10 Related work and scientific positioning

## 10.1 Failure attribution and intervention-driven debugging

Who&When formalizes responsible-agent and decisive-step attribution [15]. TRAJDEBUG traces error lifecycles to distinguish local errors from failures that remain consequential [14]. These targets concern error localization and propagation; ours concerns which reconstructed contexts qualify for later analysis, not attribution accuracy.

DoVer actively tests message or plan edits rather than relying on log-only hypotheses [13]. Causal Agent Replay models interventions and outcome distributions [17], and ReplayRCA evaluates changed actions against paired original-action placebos after restoration [2]. These are established motivations for counterfactual evaluation. We neither rerun their methods nor infer repair utility from matching observations.

Who&When Pro creates labeled failures by replaying successful prefixes before error injection and describes cached tool observations and browser replay checks [16]. AED retains source evidence and uses matched original-action retries where supported [5]. Our population is a census of archived unsuccessful code-agent endpoints. We do not claim these papers omit provenance/fidelity controls or that their implementations fail on our cases.

## 10.2 Replay fidelity, randomness, and observable agreement

Pakhomov and Nijkamp call for measured restoration agreement for the intended attribution contrast, show RNG-stream-position omissions, and report an AppWorld file-restore probe requiring cached-engine identity control [11]. These general lessons predate this manuscript. Our comparison holds 71 historical failures fixed while crossing reconstruction control with printed-output, database, and verdict targets; it also tests a distinct post-oracle export-reference path. The literal RNG probe is case-specific evidence, not discovery of the general RNG hazard or a live-versus-restored certificate.

DFAH distinguishes repeated final decisions, tool-name sets, and ordered paths while preserving execution records [12]. Different comparison levels changing apparent agreement is therefore not a new principle. We quantify how the N/C admission gap contracts from 26 to four under declared database/oracle targets, then inspect actual rows and fixed-program effects left outside those checks.

SymTrace matches canonical requests and states assumptions for represented logical state and external effects [3]. Our prefix token arguments can differ, so our witness does not falsify its stronger guarantee. Chronicle exercises changed code at replay cut points [4]; The Replay Gap switches continuation models and compares live behavior with static log stitching [6]. Here the added program is fixed and native code executes; the reconstruction/validation contrast is different.

## 10.3 Test oracles and benchmark validity

When Agent Evaluation Becomes Unreliable already studies measurement contracts, oracle integration, and an AppWorld portability boundary [7]. Its reference-solution terminal state differs from our archived agent-end reference. We localize a particular binding/view path with unchanged original cached rows and validate its export correction across all ten affected C cases. Neither oracle-integrity testing nor cache identity is introduced as new theory.

EigenData audits BFCL schemas, implementations, and reference trajectories and proposes outcome-aware evaluation [18]. CAREFUL-Bench studies effects and explicitly discusses the private current-directory omission [9]. Our natural BFCL pilot checks reproduction of archived observations and diagnostics, not semantic correctness of all task specifications; the cwd construction is a known observer calibration.

# 11 Conclusion

In 71 historical AppWorld failures, execution control changes strict admission by 26 cases, but the gap contracts to four at the checked database/oracle endpoint. Four discrepancies persist in materialized rows. A declared same-argument API call distinguishes one accepted endpoint pair through omitted RNG state; aligning future RNG removes that selected difference. A separate native program and credential-value controls isolate an authentication/write effect with retained missing repetitions. On all 10 affected C cases, native export from original retained trackers removes post-oracle comparison exclusions without changing original output contents.

Together, these observations identify a measurable trade-off: a weaker projection admits more historical executions, while leaving state required by some specified later operations unresolved. Explicit reference objects and observation stages also separate reconstruction discrepancies from validator-view discrepancies. The evidence characterizes this trade-off in a finite historical cohort; it does not establish diagnosis difficulty, repair benefit or method rankings.

An open question is which state projections are sufficient for a declared family of continuations, and when admission changes translate into changes in method utility. The present audit supplies a controlled empirical starting point while leaving those broader effects to explicitly designed comparisons.

### Artifact availability and preparation

Experiment records were frozen before this rewrite. Verified source/archive hashes and a saved runtime image preserve the previous manuscript, raw logs, unknowns and deviations. All numbers and figures here are read from that evidence; preparation ran no new task or model execution. The complete local archive contains protected benchmark-derived material and is not a public redistribution package. Aggregate tables, redacted programs, version pins and an evidence-to-claim map accompany this draft. Any later shared package must use compliant benchmark-data access rather than plaintext raw-task redistribution.